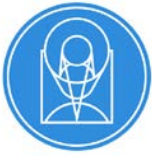

Instrument Science Report SMO 2020-01

# Recommendations for optimizing rapid ultraviolet HST observations of gravitational wave optical counterparts

Lou Strolger, Armin Rest, Ori Fox, Annalisa Calamida, Russell Ryan & Neill Reid
February 16, 2020

**ABSTRACT**

*This document presentes general guidelines to investigators proposing ultra-rapid target of opportunity (ToO) observations with the Hubble Space Telescope (HST). Establishing clear plans in advance and communicating with STScI staff, particularly the Program Coordinator, are crucial to minimising the time between triggering a ToO and executing the observations.*

## 1. Introduction

Rapid ultra-violet (UV) observations are key to understanding nature of gravitational wave (GW) sources involving neutron star progenitors, i.e., binary neutron star (BNS) and black hole-neutron star (BH-NS) mergers. Deep UV observations at early times are critical to characterizing the composition of the emission, and uniquely achievable by HST (Margutti et al, 2019). The BNS event GW170817 took the community by surprise and revealed a lack of coordination in the response. A key lesson, therefore, is that there needs to be clearer guidance on the planning and coordination of ultra-rapid ToOs if HST is to provide data in a useful timeframe.

Since the start of the LIGO/Virgo Collaboration's Observing Run 3 (O3), the community has come to appreciate that GW sources are relatively common (roughly 1 per week), but neutron star sources that can produce an accessible kilonova (ideally at distances less < 120 Mpc) are not. It is therefore important for the community to make the best use of such opportunities as they arise to maximize the overall scientific return. The goal of these notes is to provide guidance on what actions and procedures are necessary to minimize the turn-around time for HST observations.

This document builds on recommendations by the HST-LIGO Working Group (Margutti et al, 2019) for follow-up observations of gravitational-wave electro-magnetic (GW-EM) counterparts. The focus is on obtaining near-ultraviolet (NUV) spectroscopic and photometric follow-up within ~2 days of a GW event to maximize the scientific returns from an ultra-rapid trigger, taking into account the inherent difficulties involved in scheduling such observations. We emphasize the importance of maintaining clear communications with Program Coordinators (PCs), schedulers, and instrument teams at STScI. *These recommendations align with the goals of the NASA GW-EM Counterpart Task Force, assessing the contributions of NASA missions to GW-EM Astrophysics.* They are also relevant for other science programs that require ultra-rapid observations, such as early-time observations of core-collapse supernovae or gamma-ray bursts.

## 2. Scheduling observations on HST

The HST observing schedule is created through a two-stage process, largely based on systems developed in the 1990s. The first stage maps the potential observing windows for individual HST observations over the full annual cycle, developing the Long Range Plan (LRP). The second stage draws observations from the LRP to create the detailed HST observing schedule on a week-by-week basis. The observing schedule and command loads are developed by STScI staff, working in co-operation with NASA Goddard Space Flight Center (GSFC) staff who are responsible for verifying and uploading the commands to Hubble. Both STScI and GSFC staff support Hubble on a 5-day, 9am to 5 pm basis for routine operations.

HST observing programs are accepted for execution through the appropriate review process, primarily the annual meeting of the Telescope Allocation Committee. Successful teams submit a Phase II proposal that provides a detailed description of the observations, including any orientation or timing constraints. The latter constraints are combined with the object visibility to identify the possible windows for each observation.

The HST scheduling group use the SPIKE planning tool (Johnston & Miller, 1994) to combine the individual planning windows to build the LRP, mapping the range of *potential* observing opportunities for finalized programs over the full annual cycle. The LRP is a dynamic quantity, updated regularly as observations are scheduled and executed on HST, and as new observations are added through DD proposals, mid-cycles calls or joint proposals allocated by the Chandra, XMM or NRAO TACs.

The detailed HST observing schedule is developed on a week-by-week basis. HST has a 96-minute orbit, corresponding to 105 physical orbits per week. HST does not carry out observations while passing through the South Atlantic Anomaly (SAA), which occurs 7 or 8 times each day. The impact can be mitigated by selecting targets that are occulted by Earth during the SAA passage ("SAA hiders"), but such opportunities are not always available. Partial orbits can be used for shorter SNAPshot observations.

The spacecraft and science instrument commands for a given week's observations are uploaded to the telescope as a Science Mission Specification (SMS) on Sunday evenings, generally set to start executing at 0 hours UT on Monday. The timeline for preparing and finalizing the SMS is as follows (see also MacKenty, 2019):

- T-12 (Wednesday): The appropriate pool of visits from the LRP is identified and compiled for the upcoming SMS.

- T-11 & 10 (Thursday/Friday): The Science Planning and Scheduling System (SPSS) is used to build an observing calendar, optimizing the observing efficiency while taking into account specific constraints for time-critical observations. Under typical 3-gyro operations, 80-84 orbits are scheduled for GO science programs together with 10-15 SNAPs. The calendar includes high-level schedule items such as start/end slews, guide-star acquisitions, small angle manoeuvres and exposures. The Science Commanding System (SCS) is used to generate the baseline SMS.

- T-7 (Monday): Tracking and Data Relay Satellite System (TDRSS) services are requested to support the necessary command uplink and engineering and science data downloading. Contacts are added to the schedule, which may result in modifications to the calendar. There are two TDRSS opportunities per orbit and HST typically utilizes 100-120 contact opportunities in a given week.

- T-6 to 4 (Tuesday through Thursday): The POCC (Payload Operations Control Center) Analysis Support System (PASS) identifies and merges the necessary spacecraft commands with the SMS and generates the command loads and necessary ancillary products.

- T-4 & 3 (Thursday & Friday): SMS is sent to GSFC for final command load verification and processing for uplink to HST; work starts on building the subsequent SMS at STScI.

- T-1 (Sunday night local – Monday 0 hours UT): SMS begins execution.

The overall timeline is illustrated in Figure 1; the last observation in a given SMS executed 18 days after the building process started.

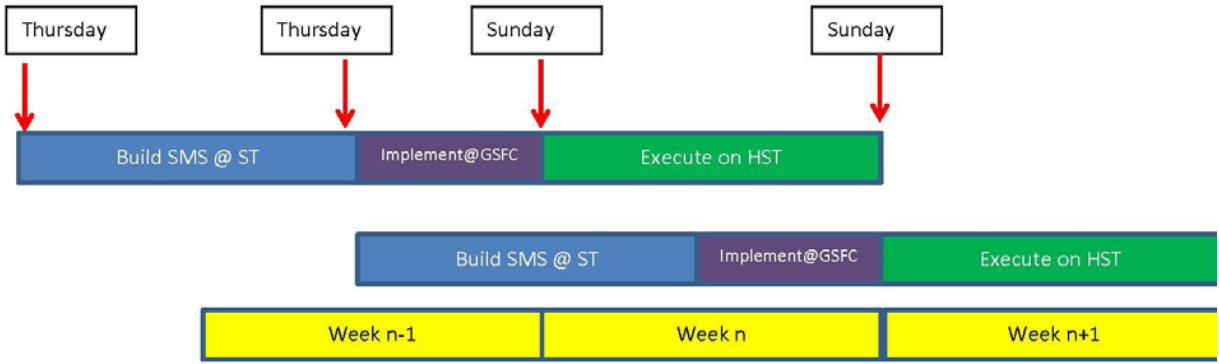

Figure 1: Timeline for constructing the HST observing schedule

## 3. Turn-around time and the coordination of transient observations

Figure 1 presents the ideal case for scheduling observations of sources and phenomena within a predictable, static universe. Observations of transient phenomena need to be incorporated into this schema, either as foreseen events through Target of Opportunity (ToO) programs, or as unforeseen events observed via Director's Discretionary Time (DDT) allocations. The incorporation is straightforward in the case of phenomena that change on a relative long time-scale: for HST, programs that can accommodate a response time longer than 21 days are regarded as non-disruptive, since the observations can be incorporated seamlessly within the build process outlined in the section 2.

Observations that require a response time for HST of less than 21 days are defined as disruptive, since they generally require modification to the SMS that is currently under construction. Even at the earliest stages of construction, incorporating new observations in an SMS is rarely as simple as a one-for-one swap; in general, maintaining observing efficiency requires significant effort on the part of the planning and scheduling team to identify and implement the optimal rebuild of the SMS. In the event of a rapid turnaround observation, the SMS incorporating the new observations needs to be built (by STScI staff), the current SMS interrupted and the new SMS uploaded to HST for execution (by GSFC staff).

In targeting GW-EM counterparts, the goal is to schedule follow-up observations as rapidly as possible. For HST, this corresponds to an ultra-rapid observation, where the effective *turn-around time*, measured from the time of submission of a Phase II observing plan via APT (or the ToO trigger of an existing Phase II proposal) to the first exposure, is less than 48 hours. Achieving this goal can be complicated by several factors, including the existence of important and/or time-critical observations that must be executed in the same timeframe, or whether the observation is triggered outside normal working hours, particularly if submitted during a weekend or holiday. **Preparation** and **communication** are key to mitigating the impact of those factors, as described in more detail in the following section.

The instruments on HST have constraints that limit their use for ultra-rapid turnaround observations. The Far-UV instruments, the Cosmic Origins Spectrograph (COS), Space Telescope Imaging Spectrograph (STIS/MAMA), and Advanced Camera for Surveys Solar Blind Channel (ACS/SBC) have photon-counting detectors that are susceptible to damage from over-counting and are therefore subject to bright-object checks. These modes are not currently available for ultra-rapid observations. However Near-UV modes on STIS/CCD, Wide Field Camera 3

(WFC3/UVIS), and ACS/WFC are available for ultra-rapid observations.

## 4. Target selection and triggering ToOs

GW sources present a peculiar challenge for HST observations. On the one hand, the latency of implementing HST observations argues for activating the ToO as soon as possible; on the other, the small field of view of HST makes it essential that a plausible optical counterpart is identified before the program is triggered. This makes it likely that decisions on the appropriate follow-up observations will likely have to be made as that EM counterpart is being confirmed and before it is fully characterized. We recommend that teams formulate a decision tree in advance, setting clear criteria for triggering on a potential candidate. *That plan should be shared with STScI staff, notably the PC, to ensure that they are aware of the decision process.*

The LIGO/Virgo collaboration provides distance, location, and type estimates (with accompanying uncertainties) that are fully refined within an hour or two of detection. The global reconnaissance effort that ensues will search until plausible EM candidates are identified. The full longitudinal coverage of the observatories will ensure that if there is an optical transient (OT) to be discovered, it will likely be located and imaged within 6 hours of the event detection. However, there is also a significant likelihood of confusing more common core-collapse supernovae for potential GW-kilonova OTs (see Kasliwal et al. 2019, and subsequent GCN circulars), thereby potentially wasting an ultra-rapid ToO (and associated orbits) if triggered too soon.

To mitigate this potential for confusion, yet maximize the opportunity for earliest observations, one might estimate the likelihood for confusion based on information such as the known core-collapse supernova rate in nearby galaxies, and the number of galaxies contained in the volume subtended by the distance estimate and the localization region.

Investigators must specify trigger criteria in the Phase I scientific justification assessed by the HST Telescope Allocation Committee. Once a proposal is accepted, well before its execution, the team should map out a decision tree: that is, once an appropriate target is discovered, what additional steps are required before triggering the ToO? The decisions should include specifying the level of verification that is required for an EM candidate before triggering the ToO, together with the type and cadence of follow-up observations that will be submitted. Teams should iterate with appropriate STScI staff before finalizing the decision tree to ensure its feasibility.

Once an appropriate event occurs, the team should contact the PC as soon as possible to give them warning that program activation is possible. They should also specify a cut-off time for aborting any such attempt if no suitable GW-EM candidate has been identified. This provides guidance to ensure that staff at STScI and GSFC will be available at the appropriate times. If a suitable optical candidate is identified, the team should trigger the ToO activation, submitting a Phase II APT as soon as possible thereafter.

Figure 2 illustrates a possible chain of communication for a GW-Kilonova event, and a schematic of activity for the parties involved in executing a potential ultra-rapid ToO. The submission of the Phase II HST observations via APT initiates the STScI work on the SMS revision. STScI staff will process the visits, including identifying guide stars, and incorporate them in the revised SMS for uploading to HST. The turnaround time from submission of the Phase II to on-sky observations will be at least 24 hours and is more likely to be closer to 36 hours.

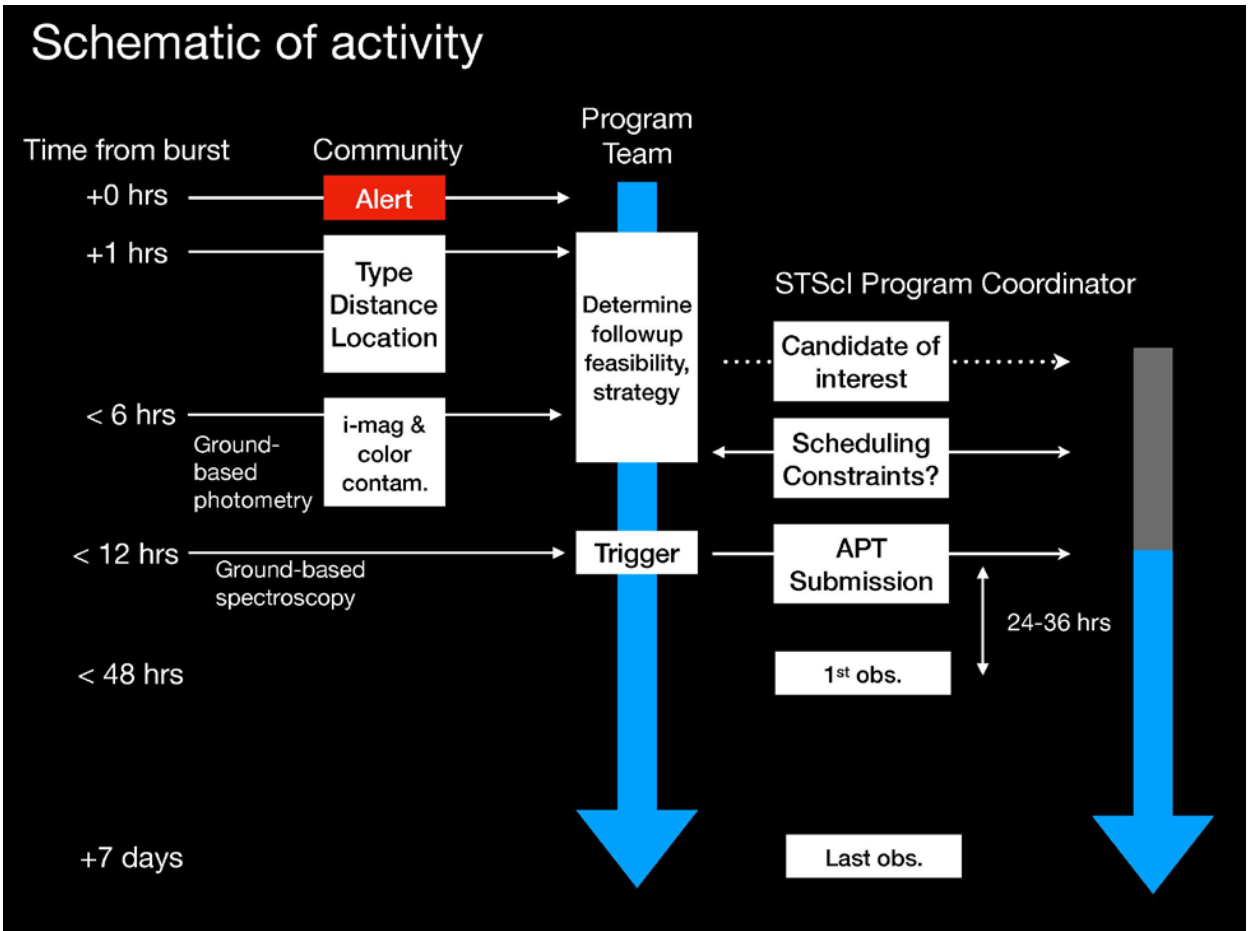

Figure 2: Schematic outlining options for the flow of communications among the community, the investigators and the schedulers at STScI. It is vital to establish early communications with the STScI PC to ensure timely scheduling even if the exact set of observations has not yet been determined.

## 5. Observation planning

Besides setting a clear trigger decision tree, teams should also devise several observing plans that are robust to likely contingencies. Teams can use APT to develop a series of templates well in advance of any GW-EM trigger. It is crucial that those templates are in place so that they can be adapted to submit the appropriate Phase II proposal as soon as possible.

Changes to ToO observing plans have the same constraints on turn-around time (2-5 days) as the initial trigger itself and are discouraged in the strongest terms. Investigators should develop observing plans for the Phase II submission that carry through well after the initial visit has been completed, extending at least one week after event detection (depending on the time allotted to the program)[1]. In addition, while it generally takes about one day for observations to be made available in the MAST archive[2], some observations may be delayed for several days before they are available. It is therefore unlikely that it will be feasible to "preview" data, or use the first HST observations to inform decisions on last sets of observations before the source has faded beyond detection.

It is also recommended that investigators devise a number of follow-up scenarios with APT, varying intended instrument/mode (spectroscopy and/or photometry), exposure times, and number of visits and cadences (spacing between visits) to suit different scenarios for the OT, based on its brightness, color or other likely characteristics. This will significantly increase the efficiency in decision making and ensure the promptest response by the observatory. Table 1 gives possible examples.

| Epoch | Phase | NUV | | | Optical | | | | NIR | | | Orbits |
|---|---|---|---|---|---|---|---|---|---|---|---|---|
| | | G280 | F225W | F275W | F336W | F438W | F606W | F814W | F110W | F160W | G141 | |
| 1 | 2 | 900 | 940 | 240 | 12 | 12 | 12 | 12 | 6 | 6 | 60 | 2 |
| 2 | 2.5 | 2400 | 2600 | 400 | 24 | 20 | 12 | 12 | 6 | 6 | 60 | 3 |
| 3 | 3 | 2700 | . . . | 700 | 40 | 30 | 12 | 12 | 6 | 6 | 70 | 2 |
| 4 | 4 | . . . | . . . | 1550 | 70 | 40 | 12 | 12 | 6 | 6 | 70 | 1 |
| 5 | 5 | . . . | . . . | 4100 | 120 | 50 | 16 | 14 | 10 | 30 | 70 | 2 |
| 6 | 7 | . . . | . . . | . . . | 700 | 210 | 30 | 30 | 35 | 35 | 400 | 1 |
| 7 | 10 | . . . | . . . | . . . | 2600 | 360 | 30 | 30 | 35 | 45 | 500 | 2 |
| Total | | | | | | | | | | | | 13 |
| GW170817-like KN at 80 Mpc | | | | | | | | | | | | |
| 1 | 2 | 2200 | 1500 | 600 | 40 | 30 | 12 | 12 | 6 | 6 | 60 | 2 |
| 2 | 2.5 | . . . | 4300 | 900 | 60 | 50 | 12 | 12 | 6 | 6 | 60 | 2 |
| 3 | 3 | . . . | . . . | 2100 | 100 | 70 | 12 | 12 | 6 | 6 | 70 | 2 |
| 4 | 4 | . . . | . . . | 3000 | 300 | 200 | 45 | 40 | 10 | 12 | 180 | 2 |
| 5 | 5 | . . . | . . . | . . . | 700 | 290 | 45 | 40 | 10 | 30 | 180 | 1 |
| 6 | 7 | . . . | . . . | . . . | 1500 | 510 | 45 | 40 | 50 | 65 | 670 | 2 |
| 7 | 10 | . . . | . . . | . . . | . . . | 1200 | 55 | 45 | 60 | 160 | 2000 | 2 |
| Total | | | | | | | | | | | | 13 |

Table 1: Nominal observing plans for a GW 170817-like kilonova at 40 and 80 Mpc with WFC3. The phase is given in days, post-merger, and the exposure times are in seconds.

[1] Investigators are expected to monitor an event and cancel any future observations that would be predicted to result in no useful data.

[2] Proposers can set scientific requirements for data receipt within days after execution (e.g. to provide pointing corrections for tightly scheduled visits); resource-intensive methods can expedite delivery of data. Such requirements must be stated in the Phase I proposal so that the resource needs can be determined and reviewed.

Scheduling individual visits can be made unnecessarily complicated by setting too restrictive a constraint on when to schedule successive visits. Each SMS includes a number of high priority observations of other targets, including time-critical observations of sources such as exoplanets. Collisions between those observations and GW-EM follow-up will require case-by-case policy decision; the STScI Director has the authority to adjudicate direct conflicts. Assigning the appropriate flexibility to GW-EM visits minimizes the potential for time-wasting iterations with the proposal team, and therefore increases the efficiency of finalizing, uploading and executing the revised SMS.

The most effective strategy for communicating the appropriate timing constraints is to write information about the approximate time of execution of each visit in the comment box provided by APT. For example, the first visit of the sequence might be specified as: "First epoch to be observed as soon as possible after the trigger (about 24 - 36 hours)". For every subsequent visit, similar comments should be added to constrain the approximate spacing between them and the first visit, which should be used as a reference. In this way, if one visit fails to execute, the link for the subsequent visits will not be lost. The second visit comment box might be, "Second epoch to be observed approximately 12±3 hours after the first epoch"; the third visit comment box, "Third epoch to be observed approximately 24±6 hours after the first epoch". Those comments provide the schedulers with both the desired observing cadence and the flexibility in each visit, allowing them to implement the program efficiently with the established time constraints. Proposers should *not* set special timing constraints within APT since those commands do not provide sufficient information on the scheduling flexibility.

## 6. Summary

This ISR describes the processes used to develop the weekly observing schedule for the Hubble Space Telescope and provides suggested guidelines that will allow users to optimize the turnaround time for ultra-rapid (< 2 day) observations of transient sources, notably optical counterparts of gravitational wave events.

In brief, the recommendations are as follows:

- Proposers should establish, well in advance, a decision tree for triggering Target of Opportunity proposals, including a clear cut-off point for activating the observations. That decision tree should be shared with the Program Coordinator at STScI.

- Proposers should prepare multiple Phase II templates to minimize the time required to finalise the actual observations.

- Wherever possible, proposers should give the PC advance warning if they believe it is likely that they will trigger their ToO on a particular event.

- When proposers trigger a ToO, they should submit a Phase II proposal that includes multiple observations (if appropriate), particularly all those scheduled in the same SMS as the initial observation.

- The timing schedule for the observations should be clearly specified with respect to the epoch of the first observation.
- The timing schedule should include the appropriate level of flexibility to minimize iterations with the STScI schedulers.
- Throughout, the proposers should maintain clear, active communication channels with STScI staff, particularly the PC.

*Acknowledgements*

We thank Denise Taylor, John Mackenty Bill Workman and Tom Brown for useful discussions and comments on the HST scheduling process and procedures.